\documentclass[amsmath,amssymb,aps,prl,superscriptaddress,twocolumn,floatfix]{revtex4}
\usepackage{graphicx}
\usepackage{dcolumn}
\usepackage{bm}
\usepackage{amsmath,amssymb,amsfonts}
\usepackage{float}
\usepackage{color}
\usepackage{dcolumn}
\usepackage{wasysym}
\usepackage{braket,mathtools}
\newcolumntype{N}{@{}m{0pt}@{}}

\begin{document}

\title{Moir\'e metal for catalysis}

\author{Yang Zhang}
\affiliation{Department of Physics, Massachusetts Institute of Technology, Cambridge, Massachusetts 02139, USA}

\author{Claudia Felser}
\affiliation{Max-Planck-Institut fur Chemische Physik fester Stoffe, 01187 Dresden, Germany}

\author{Liang Fu}
\affiliation{Department of Physics, Massachusetts Institute of Technology, Cambridge, Massachusetts 02139, USA}


\begin{abstract}

The search for highly efficient and low-cost catalysts based on earth abundant elements is one of the main driving forces in organic and inorganic chemistry.
In this work, we introduce the concept of moir\'e metal for hydrogen evolution reaction (HER). Using twisted NbS$_2$ as an example, we show that the spatially varying interlayer coupling leads to a corresponding variation of Gibbs free energy for hydrogen absorption at different sites on the moir\'e superlattice. Remarkably, as a HER catalyst, twisted NbS$_2$ is shown to cover the thermoneutral volcano peak, exceeding the efficiency of the current record platinum. The richness of local chemical environment in moir\'e structures provides an advantageous and versatile method to increase chemical reaction speed. 

\end{abstract}

\maketitle

Heterogeneous catalysis \cite{ertl1997handbook,schlogl2015heterogeneous} plays a key role in chemical and medical industries by enabling large-scale production and selective product formation. Currently, the production of 90\% of chemicals (by volume) is assisted by solid catalysts. 
An important chemical reaction for the supply of ``green energy" is to produce hydrogen (H$_2$) from water using thermal, solar or electric energy. 
In particular, the electrochemical generation of hydrogen molecules, known as hydrogen evolution reaction (HER), has the advantage of high reaction speed and eco-friendliness. 
The key component of electrochemical reaction system is electrocatalysts with high intrinsic reactivity and long-term stability. For the industry-level applications, the rational design of low-cost electrocatalysts based on earth abundant elements is in high demand.

The hydrogen electrocatalytic reaction speed is known to correlate with the intrinsic properties of solid catalysts. 
Various descriptors, such as d-band center, d-band charge, pH values, bond length, active site density, etc., have been introduced to predict the tendency of the electrocatalytic reactivities \cite{exner2020universal,liu2020progress,jiao2021descriptors}. Beyond these empirical descriptors, the most quantitative intrinsic figure of merit for HER catalyst is the Gibbs free energy ($\Delta G(H)$) of hydrogen absorption \cite{capon1973oxidation,greeley2006computational}, at the equilibrium potential. Up to now, the most efficient electrocatalysts for HER are Pt-group metals, as Pt and Pd have $\Delta G(H)$ close to the ideal HER volcano diagram peak $\Delta G(H)=0$ eV.

Metal catalysts rely on the surface dangling bonds as the active sites for chemical absorption. In industry, solid catalysts are often porous or dispersed on a supporting material to maximize surface area and enhance catalytic reactivity. However, for bulk materials, the active catalytic surface area-to-mass ratio is still quite limited. Recently, atomic thin van der Walls (vdW) materials have been studied for catalysis \cite{pandey2015two,pandey2017two,noh2018tuning,jaramillo2007identification,lukowski2013enhanced,chia2016electrocatalysis,liu2017self,shi2017two,jiang2019mos2,xie2021ws2} due to the maximized surface/volume ratio, potentially enabling ``all surface reaction".
The calculations of $\Delta G(H)$ of vdW materials within the basal plane show that the best catalyst
is monolayer TMD NbS$_2$ with $\Delta G(H)=$0.31 eV \cite{noh2018tuning}, which still falls far behind Pt/Pd
in terms of reactivity. 

\begin{figure}[t]
\includegraphics[width=\columnwidth]{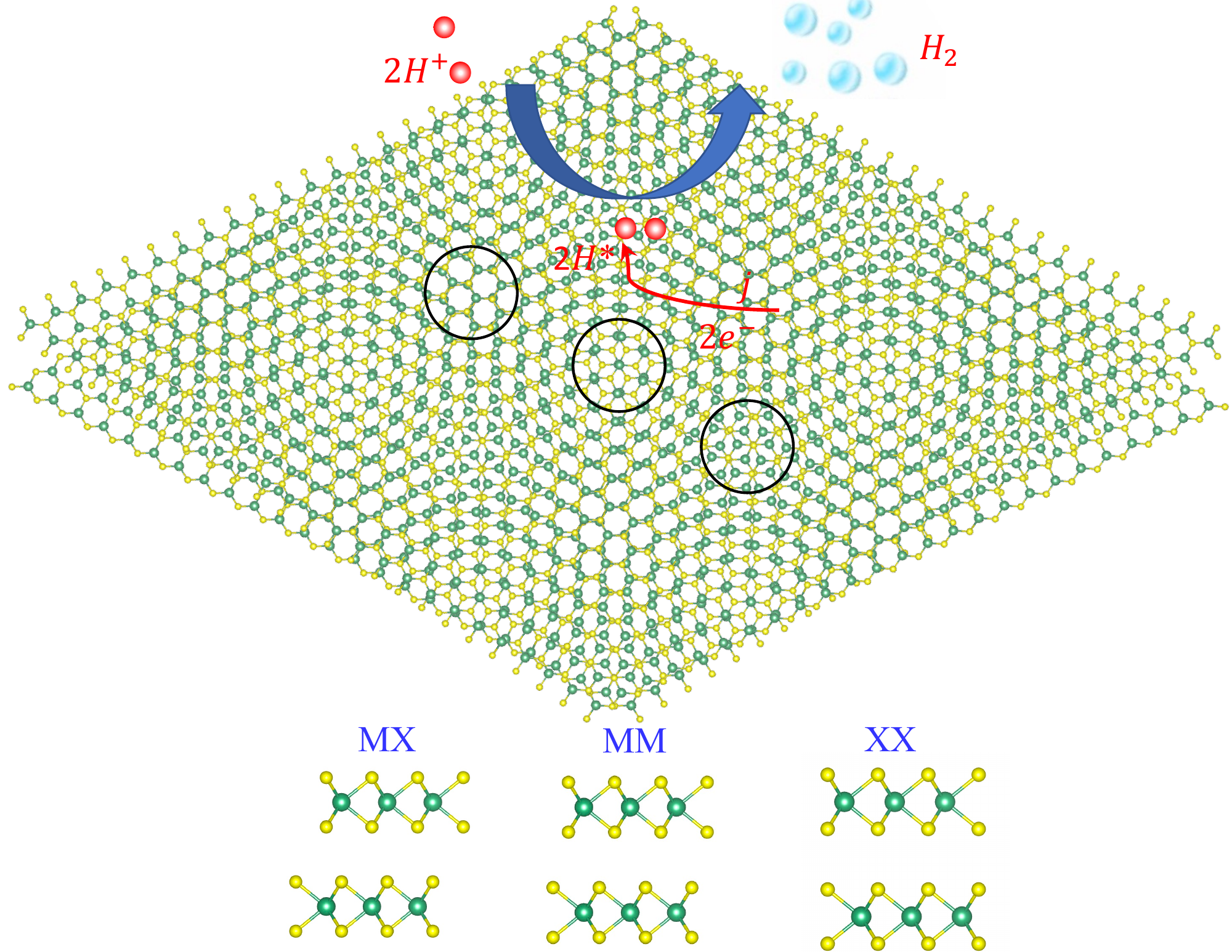}
\caption{
Schematic of HER process at three high-symmetry stacking regions of 2H TMDs, where two protons and two electrons from electrode are combined to form a hydrogen molecule. The tuning of HER catalytic properties in the bilayer vdW metals are related to the local d-band center shift \cite{jiao2021descriptors} and bandwidth at different stacking regions. We circle MX, MM and XX regions in moir\'e superlattice from left to right direction, and present the the enlarged side view of three regions at lower panel.
}\label{fig:sche}
\end{figure}

In this work, we propose a new method to enhance chemical reactivity under a broad range of environmental conditions using vdW metals with moir\'e structures. Such ``moir\'e metals" can be created by stacking two layers of identical vdW metals with a twist angle \cite{cao2018correlated,cao2018unconventional}, or two different layers with a lattice mismatch. Our key idea is that in a moir\'e superlattice, the spatial variation of stacking configuration produces a corresponding variation of Gibbs free energy for hydrogen absorption in different local regions. Such variation of $\Delta G(H)$ enables the self-adaptive optimization of chemisorption property for a wide range of reaction environments.

Using first-principles calculations, we further identify the twisted AB stacked NbS$_2$ as the superior catalyst with highest reactivity for HER compared to elementary metals and topological metals. At intermediate twist angle $\theta=5.08^{\circ}$, an 150 meV range of Gibbs free energy for hydrogen adsorption at different locations in the basal plane is observed, covering the $\Delta G(H)=0$ eV HER volcano diagram peak for standard reaction environment.

To reveal the origin of the spatial variation of $\Delta G(H)$, we calculate stacking dependent band structures and find that interlayer coupling varies significantly. By developing a simple tight-binding model, we present a heuristic argument that the spatially varying interlayer coupling on the moir\'e superlattice is responsible for the variation of $\Delta G(H)$ found in our large-scale DFT calculation. 

\begin{figure}[t]
\includegraphics[width=\columnwidth]{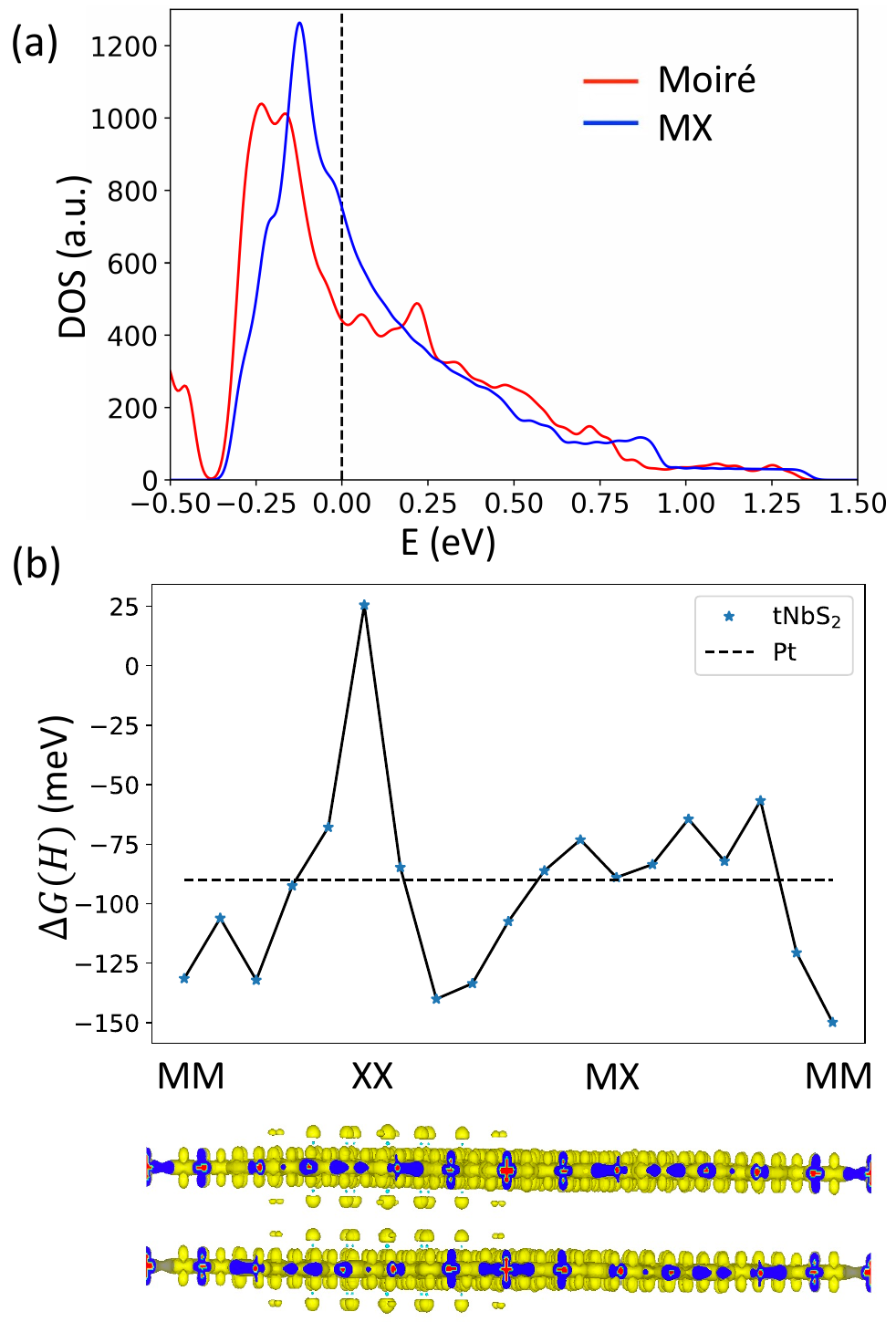}
\caption{
(a) Comparison of density of states for natural stacking MX and moir\'e superlattice. The density of states near Fermi level is significantly modified by twisting.
(b) Upper panel is the Gibbs free energy diagram for hydrogen evolution under standard conditions (1 bar of $H_2$ and pH=0 at 300 K). Energies of the intermediate states are calculated using the SCAN+rVV10 vdW functional as described in the Methods. Coverage of one hydrogen atom per moir\'e superlattice is used for all calculations. Lower panel is the wavefunction plot for twisted NbS$_2$ between $E_f-0.5$ eV to $E_f$ along MM, XX, and MX path. The wavefunction is more spread out at XX region.
}\label{fig:moire}
\end{figure}

Presently, semiconductor based moir\'e materials have been intensively studied in the physics community \cite{kennes2021moire}, such as twisted bilayer graphene \cite{cao2018correlated,cao2018unconventional} and group-VI semiconductor TMD heterostructures \cite{Tang2020,Regan2020,Shabani2021,wang2020correlated,Wu2018,Zhang2020, shimazaki2020strongly,PhysRevX.11.021027,Jin2021,Li2021,li2021quantum,Wu2019,Zhang2021,Slagle2020,devakul2021magic,padhi2021generalized,Xu2020,padhi2021generalized}. Here, for the purpose of catalysis, we consider moir\'e metals formed by stacking two layers of group-V {\it metallic} TMDs MX$_2$ in 2H structure, such as NbSe$_2$ and NbS$_2$. In the case of homobilayers with a small twist angle $\theta$ (relative to the natural bilayer stacking), the moir\'e superlattice, shown in Fig. \ref{fig:moire}, has a superlattice constant $a_{M}=a_b/\sqrt{\delta^2+\theta^2}$ where $\delta=(a_b-a_t)/a_t$, with $a_b(a_t)$ the lattice constant of each layer. In this superlattice, there are three high symmetry regions denoted as MM, XX, and MX, see Fig. \ref{fig:sche}. In MX region, the M atom on the bottom layer is locally aligned with the X atom on the top layer, as in natural TMD bilayers. On the other hand, in MM (or XX) region, the M (or X) atoms on both layers are locally aligned. Importantly, MM and XX stacking configurations are thermodynamically unstable and made possible only by moir\'e engineering. 

\begin{figure}[t]
\includegraphics[width=\columnwidth]{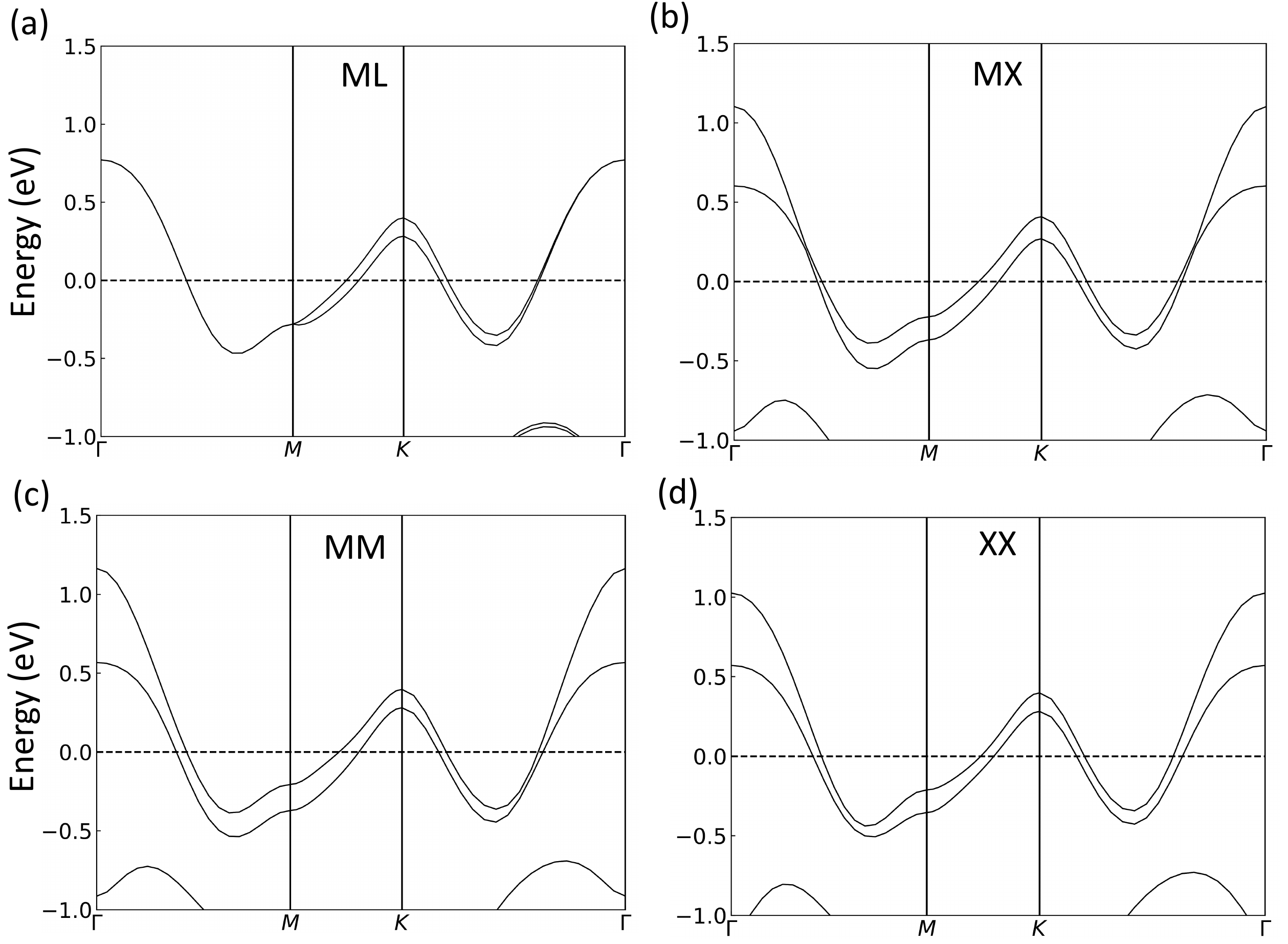}
\caption{
DFT band structure of (a) monolayer, (b) MX stacked bilayer (natural stacking), (c) MM stacked bilayer, (d) XX stacked NbS$_2$ with anti-parallel orientation. The interlayer coupling has a large variation at different stacking regions as seen in the corresponding $\Gamma$ pockets splitting.
}\label{fig:band}
\end{figure}

We calculate the electronic structures of bilayer NbS$_2$ for different stacking configurations. Monolayer NbS$_2$ features with three electron pockets near $K$, $K^{\prime}$ and $\Gamma$ valleys, and the half filled conduction bands are entirely formed from $d$ orbitals of Nb atom. 
As shown in Fig. \ref{fig:band}, 
the  $\Gamma$ valley in bilayers exhibits a large bonding-antibonding splitting on the order of 0.5eV due to vdW interlayer coupling. Importantly, this bilayer splitting varies significantly depending on the layer stacking: 0.6 eV for MM, 0.5 eV for MX and 0.46 eV for XX. The underlying variation in interlayer coupling is due to different distances between adjacent metallic atoms on the two layers.

In twisted NbS$_2$ moir\'e superlattice, the spatial variation of interlayer coupling in different local regions  provides a variety of local charging environments for the chemisorption of hydrogen atoms and molecules. The surface hydrogen binding energy directly determines its hydrogen production speed.
The HER process involves hydrogen absorption and desorption as shown in Fig. \ref{fig:sche}. Initially, the Volmer reaction transfers one electron to a proton to form an adsorbed hydrogen atom on the surface of catalyst ($H^++e^-+*\rightarrow H^*$, where * denotes the active catalyst and $H^*$ is the intermediate). Next, the desorption of $H_2$ can be achieved by either Tafel reaction ($2H^*\rightarrow H_2+2^*$) or Heyrovsky reaction  ($H^++e^-+H^* \rightarrow H_2+*$). 

Therefore, under the standard electrochemical reaction condition (room temperate and atmospheric pressure), the overall HER reaction rate is largely determined by the binding energy ($\Delta E_{H}$) of the intermediate $H^*$ and the density of the active sites in the catalyst. 
A strong bonding between hydrogen atom and catalyst slows down the desorption step (Heyrovsky or Tafel), while a weak bonding slows down the adsorption step (Volmer reaction). A direct descriptor for reaction speed is Gibbs free energy for hydrogen absorption, defined as $\Delta G_{H}=\Delta E_{H}+\Delta E_{\mathrm{ZPE}}-T \Delta S$, with $\Delta E_{\mathrm{ZPE}}$ and $\Delta S$ as the difference in zero-point energy and entropy between the adsorbed species and the gas molecule, respectively. Hereafter we use the value $\Delta E_{\mathrm{ZPE}}-T \Delta S =0.3$ eV for standard electrochemical reaction condition.
The ideal reaction rate is achieved at thermoneutral condition when $\Delta G_{H} \sim 0$ eV \cite{sabatier1911hydrogenations}.

\begin{figure}[ht]
\includegraphics[width=\columnwidth]{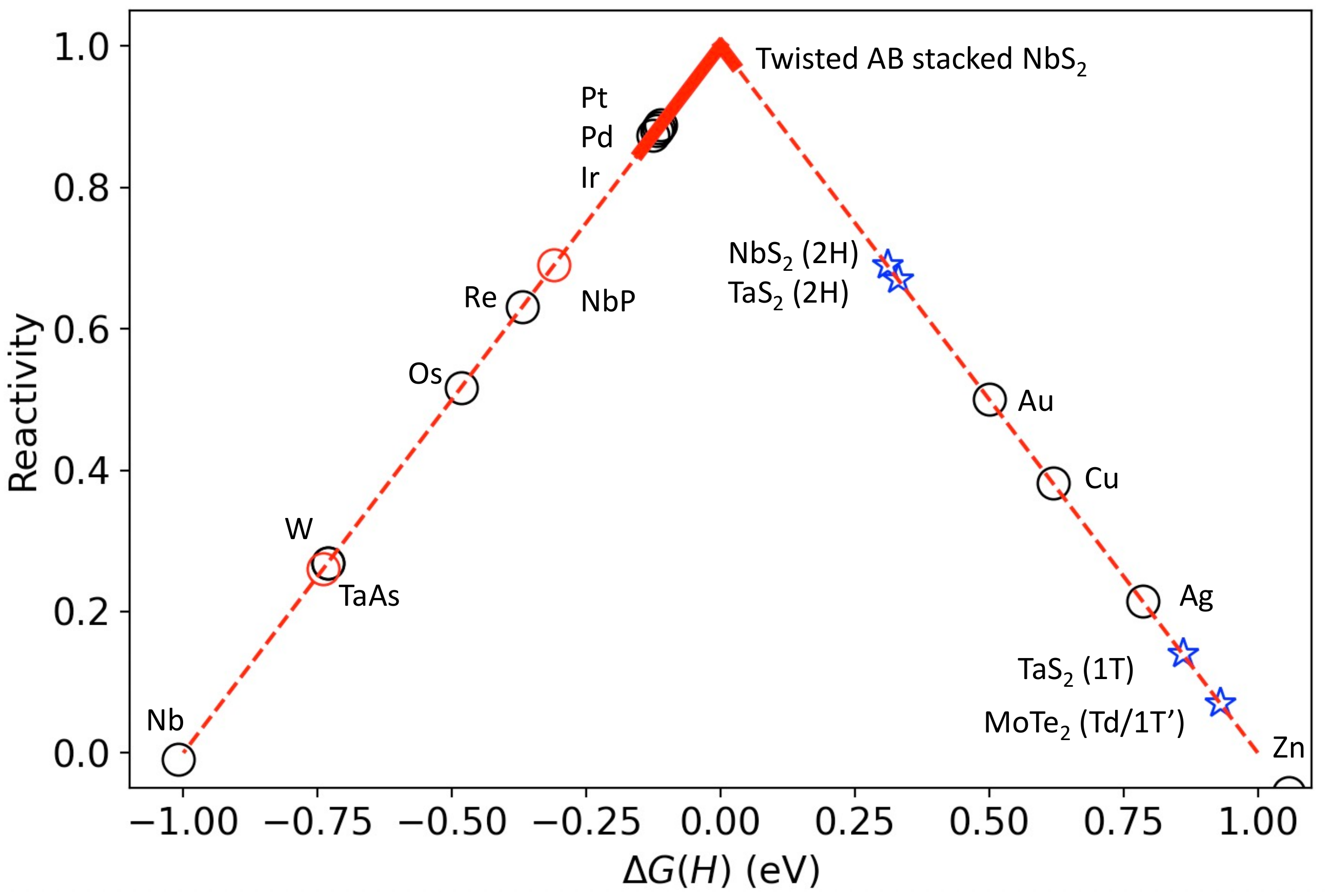}
\caption{
Predicted relative activities of various HER catalysts following the volcanic scheme as a function of calculated Gibbs free energy of hydrogen adsorption on the surface of the catalyst. Circle stands for elemental metals and topological metals, star stands for vdW metals at basal plane. The Gibbs free energy data for untwisted metal is collected from the previous literature \cite{parsons1959handbook,rajamathi2017weyl,xu2020descriptor,noh2018tuning}.
}\label{fig:gh}
\end{figure}

To investigate $\Delta G(H)$ for moir\'e metal, we first calculate the electronic structure of twisted bilayer NbS$_2$ with angle $\theta=5.08^\circ$. 
To obtain fully relaxed lattice structures, we perform large-scale density functional theory calculations with the SCAN+rVV10 van der Waals density functional~\cite{peng2016versatile}, which captures the intermediate-range van der Waals interaction through its semilocal exchange term, resulting in a better estimation of layer spacing. Owing to the different local atomic arrangements in the MM, XX and MX regions, the layer distance is found to have a strong spatial variation as shown in the supplementary material, which are 6.14\AA{}, 6.11\AA{} and 6.39\AA{} for MM, MX and XX regions in $\theta=5.08$ AB stacked moir\'e superlattices, respectively. The twisting significantly modified the density of states compared to the natural MX stacked bilayer, and the density of states peak has a shift of 100 meV. Plotting the charge density for states from $E_f-0.5$ eV to $E_f$ ( $E_f$ is the Fermi level), we find the out of plane distributions are quite different at three moir\'e regions, and XX region is more spread out than others.

In the twisted AB stacked NbS$_2$ with angle $\theta=5.08^\circ$, we calculate Gibbs free energy for hydrogen adsorption at the basal plane, especially for the line connecting three high-symmetry stacking regions as shown in Fig. \ref{fig:moire}a. $\Delta G(H)$ is obtained as $ \Delta G_{H}=\Delta E_{H}+\Delta E_{\mathrm{ZPE}}-T \Delta S$.
Here, $\Delta E_{H}$ is the hydrogen binding energy which was calculated as $\Delta E_{H}=E($ moir\'e $+\mathrm{H})-E($ moir\'e $)-1/2 E\left(\mathrm{H}_{2}\right)$ for one hydrogen atom per moir\'e superlattice. The $E($ moir\'e $+\mathrm{H})$, $E($ moir\'e) and $E\left(\mathrm{H}_{2}\right)$ are the total energy of respective parts. 

Remarkably, the spatially dependent $\Delta G(H)$ in the basal plane covers the $\Delta G(H)=0$ eV. At XX region, which is thermodynamically inaccessible without twisting, we find a Gibbs free energy $\Delta G(H)=0.01$ eV. At MM region with largest interlayer coupling, the $\Delta G(H)$ is -0.13 eV, lower than the natural stacking region MX with $\Delta G(H)=-0.09$ eV. In the line plot across three high-symmetry stacking regions as shown in Fig. \ref{fig:moire}d, we find half of the basal plane sites have absolute reactivity comparable or better than platinum.

In the volcano-shaped diagram shown in Fig. \ref{fig:gh}, we collect the Gibbs free energy for hydrogen adsorption of the twisted vdW metal NbS$_2$ with a variety of single metals, topological metals and 2D materials collected from the literature \cite{trasatti1972work,capon1973oxidation}. Among all studied bulk materials for HER catalysis, Pt with a Gibbs free energy $\Delta G(H)=-0.09$ eV is the most efficient electrocatalyst thus far, and used in industry despite the high material cost. A recent experiment on doped bulk NbS$_2$ demonstrates the ultrahigh-current-density for hydrogen evolution, which is close to platinum \cite{yang2019ultrahigh}.
Unlike the limited types of catalytic sites in conventional bulk or vdW materials, moir\'e engineering provides a wide and adjustable range of hydrogen absorption energies for the HER process in various environments with different pressure, temperature and hydrogen potentials. Importantly, the  AB-stacked bilayer NbS$_2$ at $\theta=5.08^{\circ}$ covers the volcano peak, providing a superior HER catalyst over all known materials at equilibrium potential.

To gain insight into the origin of the stacking-dependent hydrogen binding energy, we consider a simplified tight-binding model for a twisted bilayer metal, including nearest neighbor intralayer and interlayer hopping:
\begin{equation}
t_{xy}(\mathbf{d})=t_1\left(\frac{\mathbf{d} \cdot \hat{\mathbf{z}}}{d}\right)^{2}; \\
t_z(\mathbf{d})=t_2\left(\frac{1-\mathbf{d} \cdot \hat{\mathbf{z}}}{d}\right)^{2}
\end{equation}
Herein $t_1=t_1^{0} e^{-\left(r-a\right) / \delta_{1}}$, $t_2=t_2^{0} e^{-\left(r-d_{c}\right) / \delta_{2}}$, and $a=3.34$ \r{A}, $d=6.12$ \r{A} is the inplane lattice constant and layer distance of MM stacked region, intralayer hopping $t_1^0 = 0.184$ eV is fitted to match the DFT calculated density of states of $\Gamma$ pocket, $t_2^0 = 0.30$ eV is determined by the band splitting at $\Gamma$ of MM stacking configuration, the scaling parameters $\delta_{1}=0.3$ \r{A} and $\delta_{2}=1.1$ \r{A} are determined from total energy of tight binding model.

To calculate the total energy for hydrogen atoms adsorbed at the basal plane, we introduce one hydrogen atom attached to the sulfur site with constant onsite potential and hopping energy. The onsite energy of hydrogen atom is set to be $V_{H}=0.45$ eV relative to that of Nb atoms. We fit the hopping from hydrogen to Nb atom as $t_H = 2.45$ eV by matching the DFT calculated density of states ($\Gamma$ pocket Fermi surface) in MM stacked region. Although our model is greatly simplified, it correctly captures the binding energy variation across the high-symmetry stacking regions: MM, MX and XX. We plot the Gibbs free energy in the same manner as the large-scale DFT calculation shown in Fig. S2. The tight-binding description captures the overall energy scale quite well, and shows a good agreement for enlarged $\Delta G(H)$ of XX and MM region compared to MX region, which are -96 meV, -50 meV and 1 meV, respectively.

The dependence of chemical adsorption on local stacking configuration in moir\'e metals can be directly probed by scanning tunneling microscopy (STM). Previous STM studies have directly imaged adsorbed hydrogen atoms on the surface of conventional metals \cite{tatarkhanov2008hydrogen}.
For moir\'e metal, we expect that the spatially varying surface-hydrogen bond strength leads to an inhomogeneous hydrogen density distribution that correlates with the local stacking variation.

While we have focused on the twisted homobilayer NbS$_2$ as a prime example of moir\'e metals, similar stacking dependent chemisorption is expected in heterobilayers such as NbSe$_2$/NbS$_2$. 
The powerful chemical vapor deposition (CVD) \cite{cai2018chemical} method enables the scalable preparation of high quality and low cost 2D heterostructures, towards the real world application of moir\'e metal catalysts.
Besides HER, we also expect the moir\'e engineering for metals can optimize the chemisorption of the alkali metal ions for the design of electrode materials, and nitrogen and oxygen atoms for the catalysis of small organic molecules and oxygen reduction reaction (ORR). 
The chiral nature of moir\'e superlattices further provide a unique way towards heterogeneous asymmetric catalysis.

In conclusion, we have shown that the moir\'e engineering strongly enhances the HER reactivity of vdW metals due to the spatially varying interlayer bonding strength. We have identified twisted NbS$_2$ as the ideal catalyst for the $H_2$ evolution reaction, covering the volcano peak $\Delta G(H)=0$ eV. Our work provides a guiding principle for the design of highly efficient and low-cost HER catalysts from the emerging field of artificial moir\'e materials. In a broad sense, the moir\'e engineering of surface chemical properties provides a physical way for the rational design of superior electrocatalysts for a variety of chemical reactions under different environmental conditions, with the aim of closing the carbon-, hydrogen-, and nitrogen-cycle with renewable electricity.

\textbf{Method}
We performed the density functional calculations using generalized gradient approximation \cite{perdew1996generalized} with SCAN+rVV10 van der Waals density functional \cite{peng2016versatile}, as implemented in the Vienna Ab initio Simulation Package \cite{kresse1996efficiency}. Pseudopotentials are used to describe the electron-ion interactions.  We first construct AA and AB stacked NbS$_2$/NbS$_2$ homobilayer with vacuum spacing larger than 20 A to avoid artificial interaction between the periodic images along the $z$ direction. Dipole correction is added to the local potential in order to correct the errors introduced by the periodic boundary conditions in out of plane direction. The structure relaxation is performed with force on each atom less than 0.01 eV/A. We use Gamma-point sampling for structure relaxation and self-consistent calculations, due to the large moir\'e unit cell.

\section*{Acknowledgment}
We acknowledge Yan Sun, Guowei Li, Jinfeng Jia, Vidya Madhavan and Eva Andrei for valuable discussions and comments on the manuscript. 


\bibliography{ref}
\newpage

\newpage
\begin{widetext}
\begin{center}
\textbf{\large Supplemental material for ``Moir\'e metal for catalysis"} \\[10pt]
Yang  Zhang,$^1$ Claudia Felser,$^2$ and Liang Fu$^1$ \\
\textit{$^1$ Department of Physics, Massachusetts Institute of Technology, Cambridge, Massachusetts 02139, USA} \\
\textit{$^2$ Max-Planck-Institut fur Chemische Physik fester Stoffe, 01187 Dresden, Germany}
\end{center}
\vspace{15pt}

\renewcommand{\thefigure}{S\arabic{figure}}
\renewcommand{\thesection}{S\arabic{section}}
\renewcommand{\theequation}{E\arabic{equation}}
\setcounter{figure}{0}
\setcounter{section}{0}
\setcounter{equation}{0}

\section{Gibbs free energy for hydrogen absorption from tight-binding model}
We use the tight-binding model developed in main text to evaluate Gibbs free energy for hydrogen absorption.  We calculate the total energy of moir\'e superlattice with $N$ unit cells before and after hydrogen absorption as follows: $E_{\mathrm{total}}^{\mathrm{TB}}(*)=\sum_{i \text{ occ}}^N \epsilon_{i}(*)$, $E_{\mathrm{total}}^{\mathrm{TB}}(*+H)=\sum_{i \text { occ}}^{N+1} \epsilon_{i}(*+H)$, where * stands for one moir\'e superlattice.

The hydrogen binding energy of this toy model is obtained from $\Delta E(H) = E_{\mathrm{total}}^{\mathrm{TB}}(*+H)-E_{\mathrm{total}}^{\mathrm{TB}}(*)$. Assuming standard reaction condition as in the main text, we calculate Gibbs free energy $\Delta G(H)$ and plot it along MM, XX, and MX path in Fig. \ref{fig:tb}. The overall trend matches reasonably well with DFT calculations, especially the three high symmetry stacking regions.

\begin{figure}[!ht]
\includegraphics[width=0.5\columnwidth]{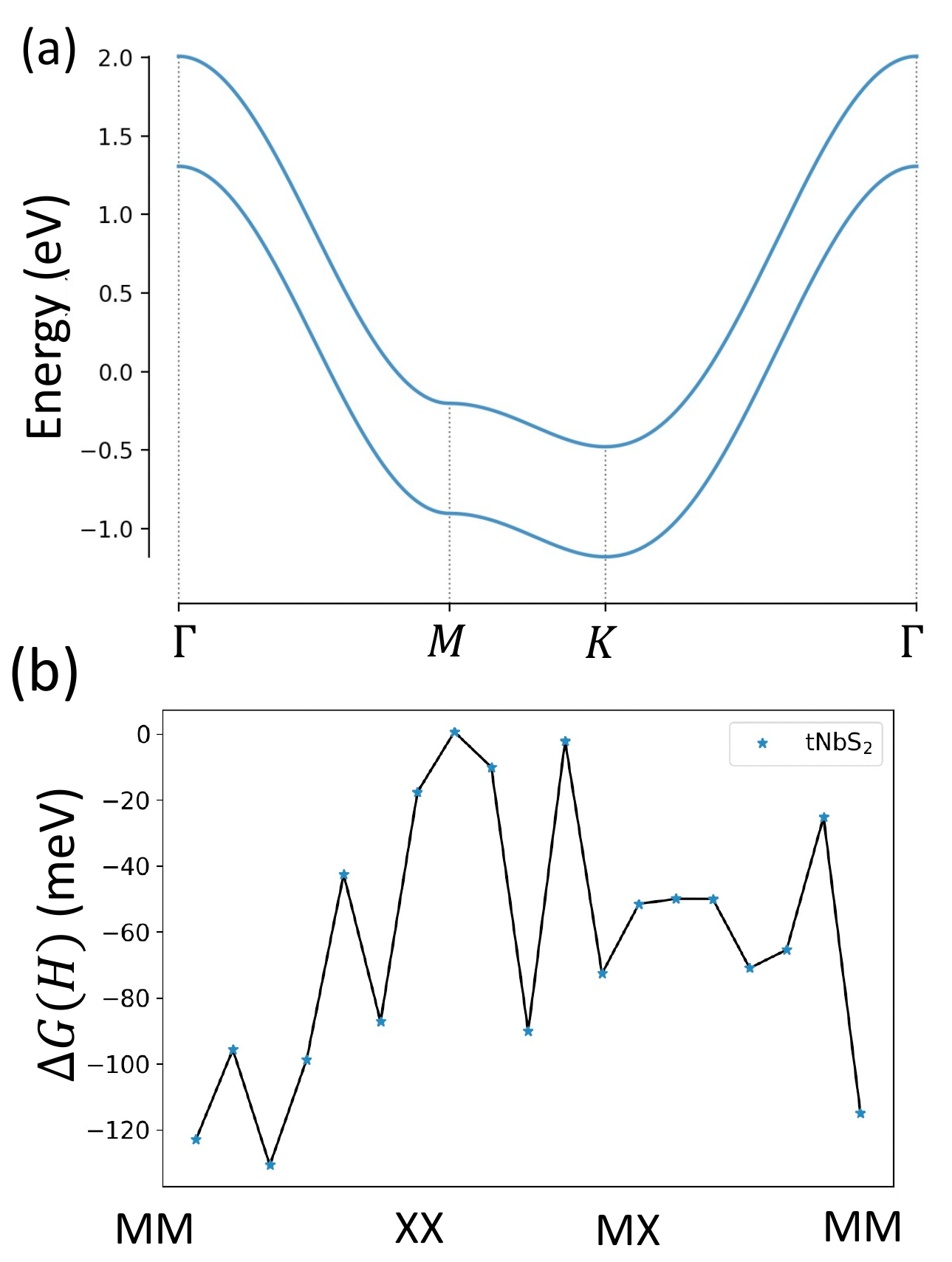}
\caption{
(a) Band structure of tight-binding model at MM stacking region,
(b) Gibbs free energy derived from hydrogen binding energy $\Delta E(H)$ calculated in tight binding model.
}\label{fig:tb}
\end{figure}

\section{Lattice relaxation of AA, AB homobilayer NbS$_2$} 
We also study TMD homobilayers with a small twist angle starting from AA stacking, where every metal (M) or chalcogen (X) atom on the top layer is aligned with the same type of atom on the bottom layer. Within a local region of a twisted bilayer, the atom configuration is identical to that of an untwisted bilayer, where one layer is laterally shifted relative to the other layer by a corresponding displacement vector ${\bm d}_0$. For this reason, the moir\'e band structures of twisted TMD bilayers can be constructed from a family of untwisted bilayers at various ${\bm d}_0$, all having $1\times 1$ unit cell. Our analysis thus starts from untwisted bilayers.

In particular, ${\bm d}_0=0, \left(-{\bm a}_{1}+{\bm a}_{2}\right) /3, \left({\bm a}_{1}+{\bm a}_{2}\right) /3$, where ${\bm a}_{1,2}$ is the primitive lattice vector  for untwisted bilayers, correspond to three high-symmetry stacking configurations of untwisted TMD bilayers, which we refer to as MM, XM, MX. In MM (MX) stacking, the M atom on the top layer is locally aligned with the M (X) atom on the bottom layer, likewise for XM. The bilayer structure in these stacking configurations is invariant under three-fold rotation around the $z$ axis.

\begin{figure}[t]
\includegraphics[width=0.6\columnwidth]{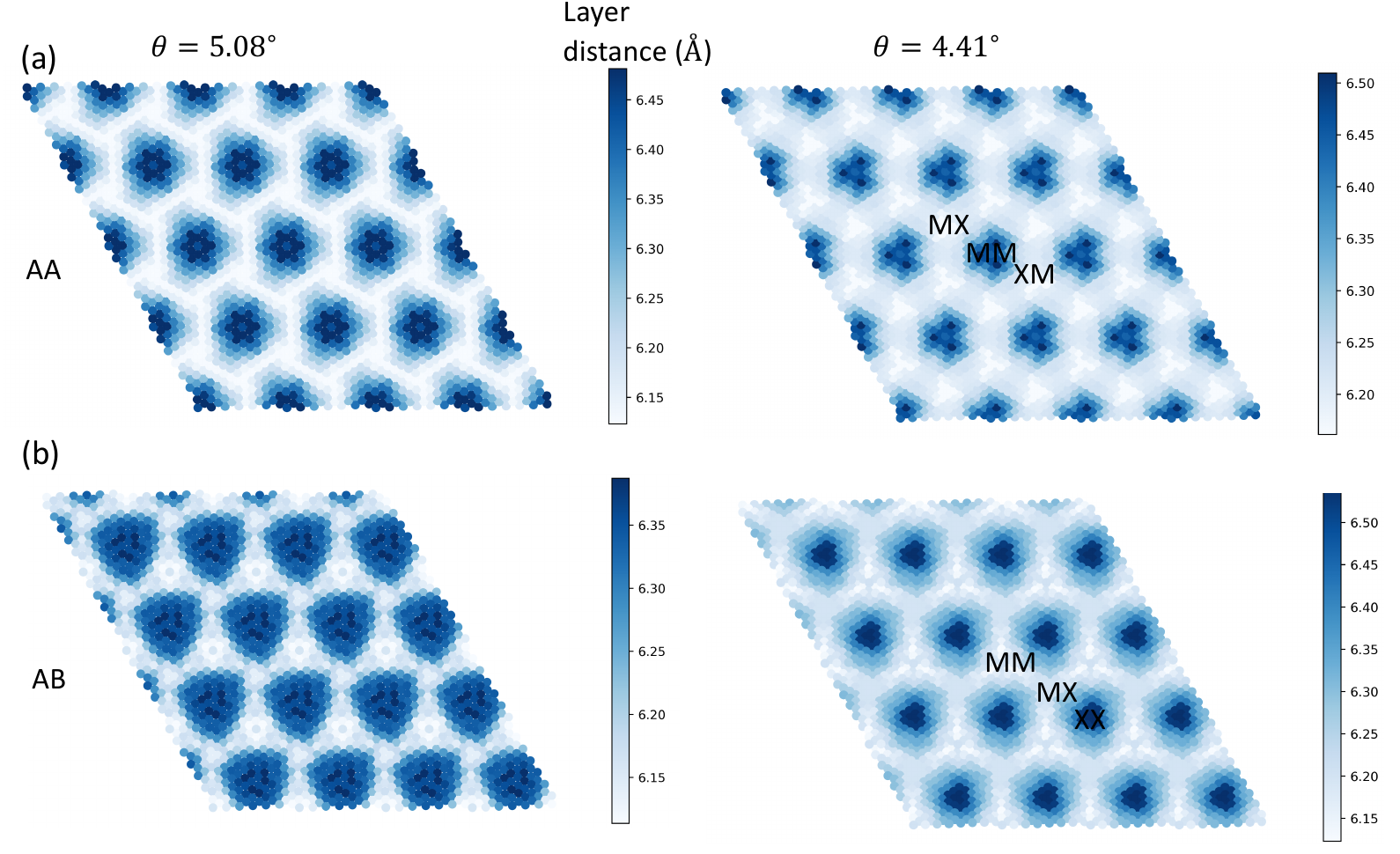}
\caption{
The interlayer distance of the twisted structure obtained from DFT is shown,
  demonstrating a large variation between different moir\'e regions. (a) AA
  stacked homobilayer NbS$_2$ with twist angle  $\theta=5.08^\circ$ and
  $\theta=4.41^\circ$; (b) AB stacked homobilayer NbS$_2$ with twist angle
  $\theta=5.08^\circ$ and $\theta=4.41^\circ$.
}\label{figS:st}
\end{figure}

We present the relaxed lattice structure and corrugation effect in AA, AB homobilayer NbS$_2$ in Fig. \ref{figS:st}.

\section{Gibbs free energy of AA stacked homobilayer NbS$_2$}
When twisting from AA stacked homobilayer, there are two high symmetry stacking regions, denoted as MM, MX, and XM region. And the MX region is symmetry related to XM region via $C_{2y}$ operation. In the untwisted structure, we see four spin and layer splitted bands at $K$ point in both MM and MX (XM) region, and an enlarged bonding-antibonding splitting at $M$ and $\Gamma$ point in MM region compared to MX (XM) region due to direct contact of metallic atom as shown in Fig. \ref{FigS:AA}. The interlayer coupling strength at MM and MX (XM) region are 0.336 eV and 0.254 eV, respectively.

\begin{figure}[t]
\includegraphics[width=0.6\columnwidth]{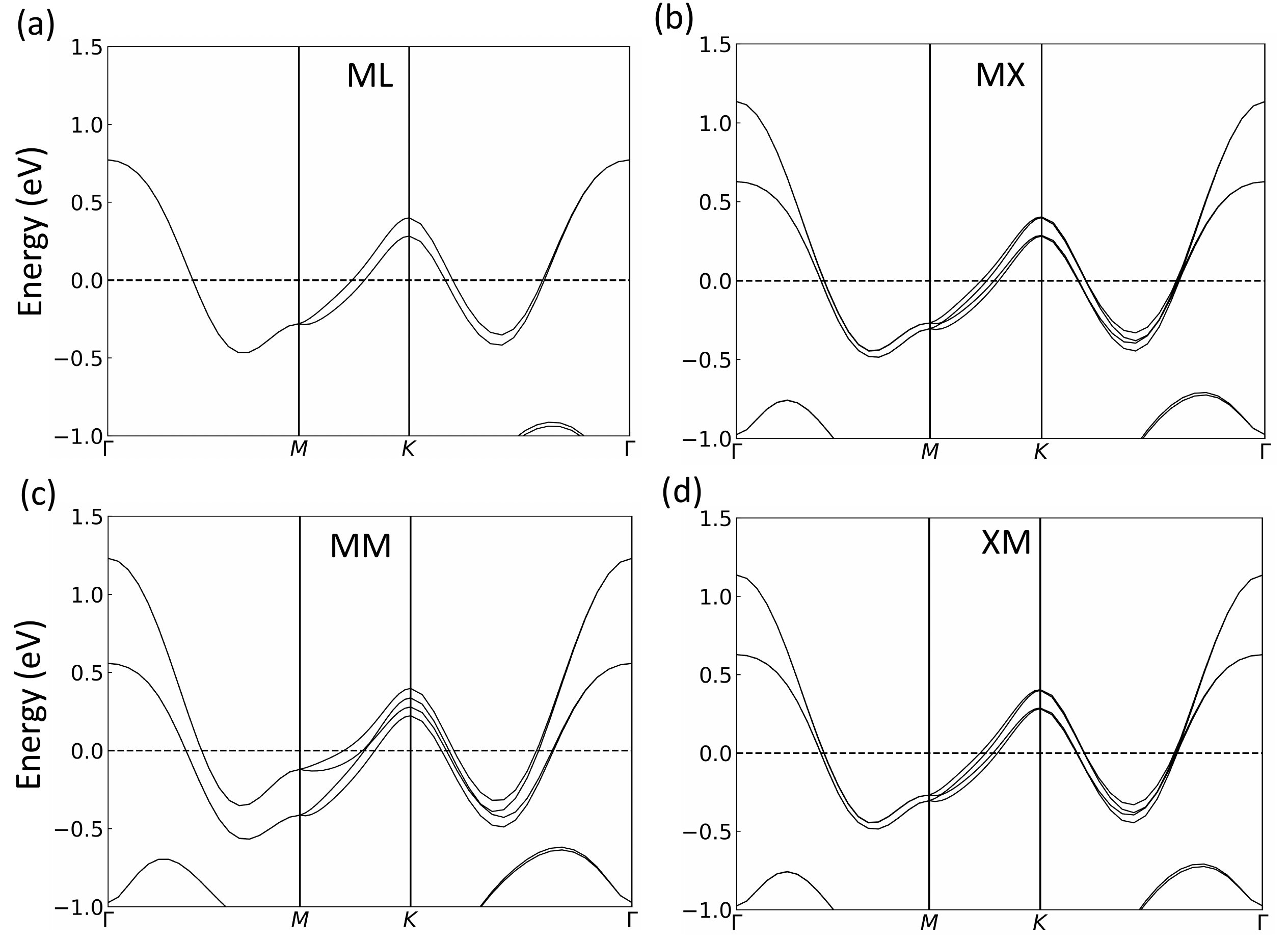}
\caption{
DFT band structure of (a) monolayer NbS$_2$; (b) MX stacked bilayer, (c) MM stacked bilayer, (d) XM stacked NbS$_2$ with parallel configuration . The interlayer coupling has a large variation at MM and MX (XM) regions as seen in the $\Gamma$ pockets splitting.
}\label{FigS:AA}
\end{figure}

We further calculate the Gibbs free energy of hydrogen absorption in AA stacked moir\'e superlattice. At twist angle $\theta=5.08$, we get -0.171 eV at MM region and -0.087 eV at MX region, with a range of 84 meV. In analogy to the AB stacking moir\'e, we plot the wavefunction from $E_f-0.6$ eV to $E_f$ to check the real space charge variation at different stacking regions.

\section{Twist angle dependent Gibbs free energy and density of states}
Apart from twist angle $\theta = 5.08^{\circ}$, we also calculated AB stacked moir\'e superlattice at nearby angle $\theta = 4.4^{\circ}$ and $\theta = 6.0^{\circ}$. For $\theta = 6.0^{\circ}$ and $L_m=3.19$ nm, we got $\Delta G(H) = -0.114, -0.046, -0.066$ eV at MM, MX, and XX region.

\section{Moir\'e band structure}
In this section, we present the band structures of moir\'e superlattices in AB and AA stacked NbS$_2$, and their density of states compared to the natural stacked bilayer. Compared to the splitted density of states peak in AB stacking moir\'e, we have a single peak in AA stacking moir\'e in Fig. \ref{figS:dos}a. The moir\'e band structures shown in Fig. \ref{figS:dos}(c,d) are massively complicated, as expected from the folding of large Fermi surface of monolayer NbS$_2$.

\begin{figure}[t]
\includegraphics[width=0.8\columnwidth]{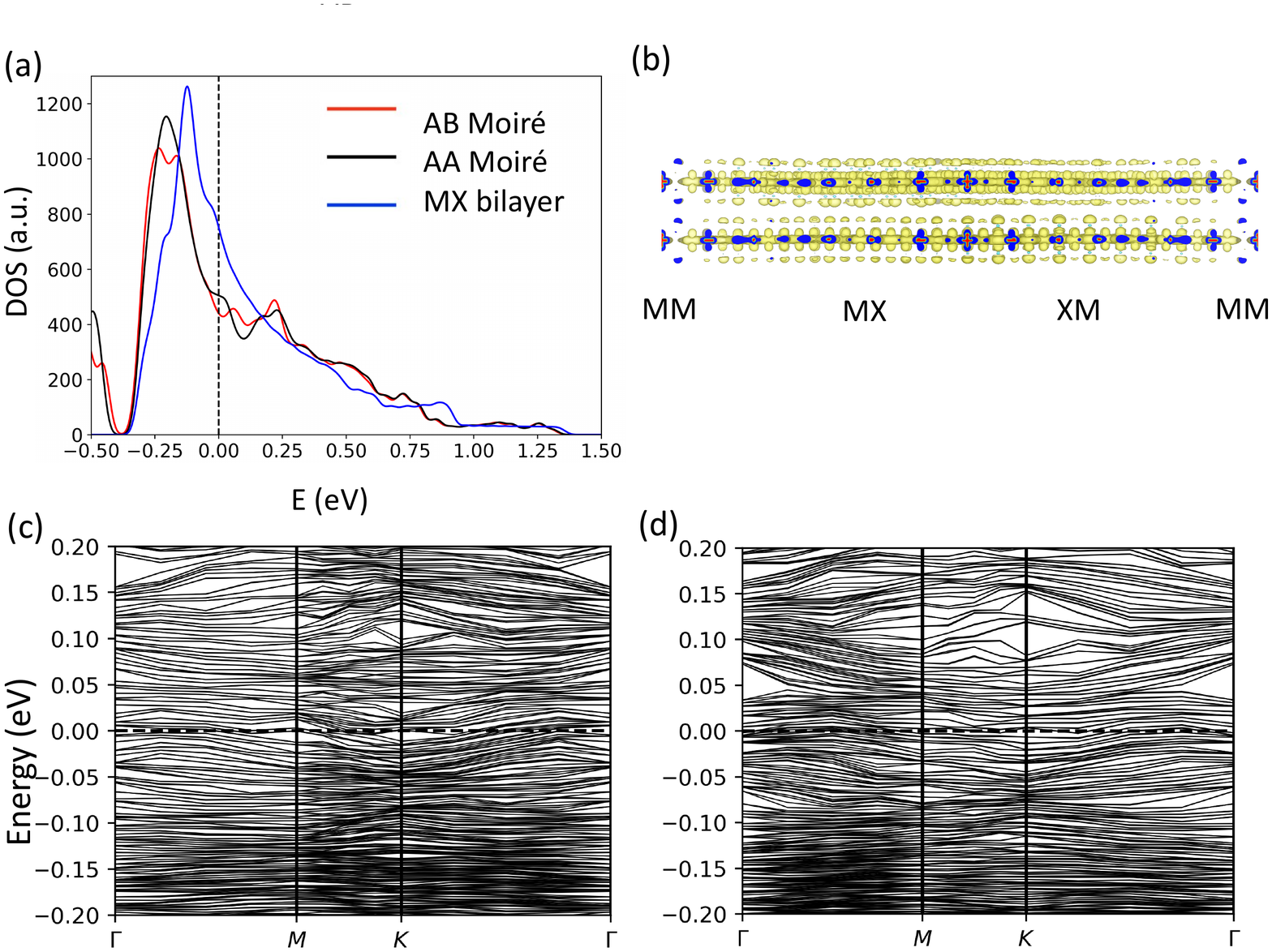}
\caption{
(a) Density of states for moir\'e superlattice with AA (AB) stacking, compared to natural MX stacking bilayer. (b) Wavefunction plot along high symmetry line for AA stacking moir\'e superlatttice of NbS$_2$ from $E_f-0.6$ eV to $E_f$. DFT band structure of (c) twisted AB stacked NbS$_2$; (d) twisted AA stacked NbS$_2$.
}\label{figS:dos}
\end{figure}

\end{widetext}

\end{document}